\documentstyle[12pt,aasms4]{article}
\newcommand{\postscript}[2]
 {\setlength{\epsfxsize}{#2\hsize}
  \centerline{\epsfbox{#1}}}
\def\tempest%
{\begin{array}{ccc}
1 & 1 & 1 \\
1 & 1 & 1 \\
4 & 3 & 8
\end{array}}

\begin{document}

\title{Where Are the Binary Source Gravitational \\
	Microlensing Events? II}
\bigskip
\bigskip

\author{Cheongho Han}
\author{Young-Jin Jeong}
\bigskip
\affil{Department of Astronomy \& Space Science, \\
       Chungbuk National University, Cheongju, Korea 361-763 \\
       cheongho@astro-3.chungbuk.ac.kr,\\
       jeongyj@astro-3.chungbuk.ac.kr}
\authoremail{cheongho@astro-3.chungbuk.ac.kr}
\authoremail{jeongyj@astro-3.chungbuk.ac.kr}
\bigskip

\begin{abstract}
The gravitational microlensing light curve deviates from the ideal 
Paczy\'nski form if either the lens or the source is composed of binaries: 
binary lens and binary source lensing events.  Currently, 6 candidate 
binary lens events have been reported, while the frequency of binary source 
events is relatively very low despite the same multiplicity of lenses and 
sources, and only a single candidate binary source event has been reported.
To account for the rarity of binary source events, Dominik pointed out that 
for a typical binary source event the separation between the component 
source stars is very large, resulting in large difference in impact parameters 
between the component binary source stars.  In these 
cases, the light curve of the highly amplified source star is barely affected 
by the light from the star with low amplification, making the observed binary
source light curve difficult to distinguish from that of a single source 
lensing event.  In this paper, we determine the fraction of events with 
similar source star amplifications is as much as $\sim 8\%$, and thus 
show that the very low detection rate for binary source events cannot be 
explained by this effect alone.  By carrying out realistic simulations of 
binary source events, we find that a significant fraction of binary source 
events are additionally missed from detection due to various other reasons.
First, if the flux ratio between the component stars is very large, 
the light curve of the bright star is hardly affected by the light
from the faint star.  Second, if the separation is too small, the binary 
source stars behave like a single star, making it difficult to separate the 
binary source event from a single event.  
Finally, although the probability of detecting binary source events increases 
as the source separation increases, still some fraction of binary source 
events will be missed because the light curves of these events will mimic 
those of single source events with longer timescales and larger values of 
the impact parameter.
\end{abstract}

\vskip25mm

\centerline{submitted to {\it The Astrophysical Journal}: Apr 27, 1998}
\centerline{Preprint: CNU-A\&SS-05/98}
\clearpage

\section{Introduction}
Experiments to detect Massive Astronomical Compact Halo Objects 
(MACHOs) by monitoring light variations of stars caused by gravitational
microlensing have been carried out and nearly 300 events have been 
detected (Alcock et al.\ 1997, Ansari et al.\ 1996, Udalski et al.\ 1997,
Alard \& Guibert 1997).\markcite{alcock1997, ansari1996, udalski1997,
alard1997}
When both the lens and source star are approximated as point sources, 
the light curve of a lensing event is related to the lens-source separation 
in units of the Einstein ring radius $r_{\rm E}$ by 
$$
A(u) = {u^2+2\over u(u^2+4)^{1/2}};\qquad
u^2 = \beta^2 + \left( {t-t_0\over t_{\rm E}}\right)^{2},
\eqno(1)
$$
where $\beta$ is the impact parameter and $t_0$ is the time of maximum 
amplification.  The Einstein ring radius crossing time 
(Einstein timescale) is related to the physical parameters of the 
lens system by
$$
t_{\rm E} = {r_{\rm E}\over v};\qquad
r_{\rm E} = \left( {4GM\over c^2}{D_{ol}D_{ls}\over D_{os}}\right)^{1/2},
\eqno(2)
$$
where $M$ is the mass of the lens, $v$ is the lens-source transverse speed,
and $D_{ol}$, $D_{ls}$, and $D_{os}$ are the separations between the observer, 
lens, and source.

The lensing light curve deviates from the ideal form in equation (1) if 
either the lens or the source is composed of binaries: binary-lens (BLL) and 
binary-source lensing (BSL) events.  Currently, 6 candidate BLL events 
have been reported by various groups (MACHO LMC\#1, Dominik \& Hirshfeld 1994, 
1996; OGLE\#7, Udalski et al.\ 1994; DUO\#2, Alard, Mao, \& Guibert \ 1995;
MACHO LMC\#9, Bennett et al.\ 1996; MACHO Bulge 95-12, Pratt et al.\ 1995;
MACHO Bulge 96-3, Stubbs et al.\ 1997).\markcite{dominik1994, 1996}
In contrast, the frequency of BSL events is relatively very low despite the 
same multiplicity of lenses and sources, and only a single candidate BSL event 
has been reported (MACHO LMC 96-2, Becker et al.\ 1997).\markcite{becker1997}
To account for the rarity of BSL events, Dominik (1998) pointed out that
for a typical BSL event the separation between the component source stars 
is very large, resulting in large difference in impact parameters between
the component binary source stars.  In these 
cases, the light curve of the highly amplified source star is barely affected 
by the light from the star with low amplification, making the observed BSL 
light curve difficult to distinguish from that of a single source lensing 
(SSL) event.  

In this paper, we determine the fraction of events with similar source
star amplifications is as much as $\sim 8\%$, and thus show that the very 
low detection rate for BSL events cannot be explained by this effect alone.
By carrying out realistic simulations of BSL events, 
we find that a significant fraction of BSL events are additionally missed 
from detection due to various other reasons.
First, if the flux ratio between the component stars is very large, 
the light curve of the bright star is hardly affected by the light
from the faint star.  Second, if the separation is too small, the binary 
source stars behave like a single star, making it difficult to separate the 
BSL event from a SSL event.  
Finally, although the probability of detecting BSL 
events increases as the source separation increases, still some fraction of 
BSL events will be missed because the light curves of these events will 
mimic those of SSL events with longer timescales and larger values of the 
impact parameter.

\section{BSL Events}

The light curve of a BSL event is the superposition of the light curves    
from the individual source stars and is represented by
$$
A_{\rm BSL} = {\sum_{i=1}^{2}F_{0,i}A(u_{i})\over \sum_{i=1}^2 F_{0,i}} 
= \sum_{i=1}^{2} {\cal F}_{i}A(u_i),
\eqno(3)
$$
where $i=1,\ 2$ denote the primary (brighter) and companion (fainter)
source stars respectively, $F_{0,i}$ are their unamplified fluxes, 
${\cal F}_i=F_{0,i}/\sum_i F_{0,i}$ are the contributions of individual 
source star fluxes to the total flux, and $A(u_i)$ are the amplifications 
of the individual source stars.  If there is no blending, 
${\cal F}_2=1-{\cal F}_1$.  The lens-source separations for the BSL event 
are related to the lensing parameters by 
$$
u_i^2 = \beta_{i}^2 + \left( {t-t_{0,i}\over t_{\rm E}} \right)^2.
\eqno(4)
$$
Therefore, compared to the 3 lensing parameters ($t_{\rm E}$, $t_{0}$, and 
$\beta$) that must be fit to a SSL event, a fit to a BSL event light curve 
requires 6 parameters (${\cal F}_1$, $t_{\rm E}$, $t_{0,i}$, and $\beta_{i}$).
However, for various reasons the light curves of BSL events are often 
difficult to distinguish from those of SSL events, resulting in the low rate 
at which BSL events are identified.  In the following sections, 
we investigate various reasons for the scarcity of detected BSL events and 
estimate what fraction of the actual number of BSL events are missed due 
to these reasons.

\section{BSL Events with Large Amplification Difference}

As one reason for the rarity of detectable BSL events, Dominik (1998) pointed 
out that if the difference in the impact parameters between the two source
stars $\delta\beta = \left\vert\beta_1-\beta_2\right\vert$ is too large, 
the amplification of one star completely dominates that of the other.
As a result, the light curve of the BSL event appears to be that of a 
SSL event, i.e., $A_{\rm BSL} \sim F_j A(u_j)$, where $j$ represents the 
source star with higher amplification.  The probability that a BSL event 
has a large $\delta\beta$ increases as the separation between source stars 
$\ell$ increases, and thus BSL events are misidentified more frequently in 
cases of wide source star separations.  Therefore, if most binary source 
stars are widely separated compared to the typical size of the Einstein ring, 
a significant fraction of BSL events will be indistinguishable from SSL events.

We now show, however, that the very low detection rate for BSL events 
cannot be explained by large values of $\delta\beta$ alone. 
To determine the fraction of actual BSL events that are misidentified 
as SSL events due to large $\delta\beta$, we compute the distribution of 
BSL events with $\delta \beta$ less than a limiting value 
$\delta\beta_{\rm lim}$ as a function of binary source separation: 
$f(\delta\beta_{\rm lim},\ell)$. 
This distribution is obtained by convolving the 
distribution of BSL separations $f(\ell)$ with the mean probability 
$P(\delta\beta_{\rm lim},\ell)$ for each BSL event with $\ell$ to have 
$\delta\beta\leq \delta\beta_{\rm lim}$, i.e., 
$f(\delta\beta_{\rm lim},\ell)=f(\ell)\otimes P(\delta\beta_{\rm lim},\ell)$.

For $f(\ell)$ we adopt the distribution found by
Duquennoy \& Mayor (1991),\markcite{duquennoy1991} which is based on the 
orbital periods of 164 G dwarf samples.  They found that the distribution 
of orbital periods appeared remarkedly symmetric and could be approximated 
by a Gaussian with a mean and standard deviation given by
$\langle \log P\rangle \sim 4.8$ and $\sigma_{\log P}\sim 2.3$ in units of days.
When the typical mass of a binary system is assumed to be $\sim 1\ M_{\odot}$,
the distribution of binary source star separations has also a Gaussian form
with the mean and standard deviation given by
$\langle \log\ell \rangle \sim 1.5$ and $\sigma_{\log \ell}\sim 1.5$
in units of AU.  This form of $f(\ell)$ is shown in the top panel of Figure 1.

To compute $P(\delta\beta_{\rm lim},\ell)$, we carried out a simulation in 
which we produce a set a binaries whose component stars are separated 
uniformly from zero to infinity.  For each binary, we assign the impact 
parameter for the first event, which is uniformly distributed over the range 
$0\leq\beta_1\leq 1$.  Once $\beta_1$ is assigned, the impact parameter of 
the second star is assigned by 
$\beta_2 = \left\vert\beta_1 + \ell\sin\theta\right\vert$
and $\delta\beta$ is computed as
$\delta\beta = \left\vert\beta_1-\beta_2\right\vert$.
Here the position angle $\theta$ is randomly distributed
in the range $0\leq\theta\leq 2\pi$.
Unlike $\beta_1$, we allow $\beta_2$ to have any value.
Therefore, according to our definition of a BSL event, both stars in the 
binary do not necessarily pass inside the Einstein ring as long as 
at least one of the two stars does.  Finally, the probability 
$P(\delta\beta_{\rm lim},\ell)$ is obtained by computing the 
fraction of all events yielding $\delta\beta\leq\delta\beta_{\rm lim}$.
In the middle and lower panels of Figure 1, we present 
$P(\delta\beta_{\rm lim},\ell)$ and the corresponding distributions of 
$f(\delta\beta_{\rm lim},\ell)$ for the two cases 
$\delta\beta_{\rm lim}=0.1$ and $1.0$.  Here we have assumed 
$r_{\rm E}=1$ AU, which is the typical Einstein ring radius 
for Galactic bulge events, with $M\sim 0.25\ M_{\odot}$,
$D_{ol}\sim 5\ {\rm kpc}$, and $D_{os}\sim 8\ {\rm kpc}$, which comprise
the majority of detected microlensing events.

From the distribution $f(\delta\beta_{\rm lim},\ell)$ we find that the 
fraction of BSL events with equivalent source star impact parameters is
still substantial.  For example, the fraction of BSL events with 
$\delta\beta\leq\delta\beta_{\rm lim} = 0.1$ comprises $\sim 8.2\%$
of the total (represented by the dark shaded region in the lower 
panel of Figure 1).
Assuming a stellar duplicity rate of $\sim 50\%$ 
(Abt \& Levy 1976)\markcite{abt1976}, if large $\delta\beta$ were the
only reason for the low rate of BSL event detections, among the total of
$\sim 300$ microlensing events one would expect to find $\sim 12$ BSL events, 
far exceeding the single detection to date.  Therefore, additional reasons 
are required to explain the rarity of BSL events.

\section{BSL Events with Equivalent Amplifications}

In the previous section we showed that the very low detection rate of BSL
events cannot be explained solely by large differences in amplification
and so additional reasons are required.  To find these reasons, we carry 
out realistic simulations of BSL events whose source star amplifications 
are equivalent.  In order to match the observational conditions of 
the current lensing experiments, in the simulation, we assume 
observations are made using a 1.27 m telescope with a dichroic beam splitter 
for simultaneous imaging in two bands, following to those of the MACHO group.  
The CCD camera can detect 25 ${\rm photons}\ {\rm s}^{-1}$ from a 
$V=20$ mag.\ star on a 1 m telescope.  
We assume a magnitude of $m=17$ for the primary star, which is 
typical for Galactic bulge source stars, and that the brightness 
difference between the primary and secondary stars, $\delta m$, follows 
the distribution $f(\delta m)$.  To see how our results depend on the 
distributions $f(\delta m)$ for different spectral types of primary source 
stars, we examine 3 types: A, B, and G dwarfs.  The distributions 
$f(\delta m)$ are adopted from Abt \& Levy (1985)\markcite{abt1985} for A 
stars, Abt, Gomez, \& Levy (1990)\markcite{abt1990} for B stars, and 
Duquennoy \& Mayor (1991)\markcite{duquennoy1991} for G stars, and are 
listed in Table 1.  Note that these distributions are identical to those 
adopted by Griest \& Hu (1992)\markcite{griest1992} to determine the 
frequency of various types of BSL events.

Once both source stars are selected, we produce BSL events
according to equations (3) and (4), in which the difference in impact 
parameters is restricted to be less than 0.1.
In our example, events are assumed to have $t_{\rm E} = 15\ {\rm days}$, 
which is the most common value observed for Galactic bulge events.
The observations are assumed to be carried out twice each night during the 
period $-3t_{\rm E}\leq t\leq 3t_{\rm E}$ with the high photometric 
precision of $1\%$.
Then the light curve of each BSL event is fitted with a SSL event light
curve as given in equation (1) and the resulting $\chi^2$ is obtained by
$$
\chi^2 = \sum_{i=1}^{N_{\rm pt}} \left[ {N_{\rm SSL}(t_i)-N_{\rm BSL}(t_i) 
\over pN_{\rm SSL}(t_i)} \right]^{2},
\eqno(5)
$$
where $N_{\rm pt}$ is the number of data points, $p=1\%$ is the photometric 
precision, $N_{\rm SSL} (t)$ is the photon counts of the theoretical SSL 
event light curve, and $N_{\rm BSL}(t)$ represents the counts predicted for 
the simulated BSL event.  We define `detectable BSL events' to be the BSL 
events whose light curves can be distinguished from those of SSL events 
with a confidence level greater than $3\sigma$.

In Table 1, we list the mean probability $P_{3\sigma}(\delta m)$ 
of detecting a BSL event with $\delta m$, averaged over the entire ranges 
of $\ell$ and $\beta$.   We also list the distribution of BSL events which 
can be distinguished from SSL events at a high confidence level as a function 
of source brightness difference,  which is obtained by 
$f_{3\sigma}(\delta m) = P_{3\sigma}(\delta m) f(\delta m)$.
From the table one can find several trends.  First, regardless of the spectral 
types, the probability of BSL events whose light curves can be distinguished 
from SSL events is very low even with similar source star amplifications and 
high precision photometry.  
The probability is especially low for events with large brightness 
differences between component stars.
For example, detecting events with $\delta m\gtrsim 4.2$ is highly unlikely; 
the number of these events comprise $\sim 1/2$ of the total 
BS events, although there is some dependence on the stellar types.
Second, as the brightnesses of the source stars become equivalent,
the probability $P_{3\sigma}$ increases.  
However, even for these events, the probability of detecting BSL events is low.
We find $P_{3\sigma}(\delta m\sim 2.5)\sim 6.5\%$ and $P_{3\sigma}
(\delta m\sim 0.7)\sim 10.7\%$.
This is because binary source stars with very small separations 
($\log\ell \leq -1.0$) behave as if they were single source stars, making 
them difficult to distinguish from SSL events; these events with small 
$\delta\ell$ comprise $\sim 75\%$ of the total number of BSL events with 
$\delta\beta\leq 0.1$.
Additionally, the light curves of some fraction of BSL events with medium 
size separations ($\log\ell\sim -0.5$) mimic those of SSL events with 
lensing parameters adjusted to longer timescales and larger impact parameters.
Consequently, the fraction of detectable BSL events is very small 
($\lesssim 6\%$), even among events with similar amplifications.

\section{Summary}

The detection rate of BSL events is very low relative to the detection rate 
of BLL events despite similar duplicities of lenses and source stars.  
The scarcity of BSL event detections results from a combination of reasons.
First, if the difference in impact parameters between individual components of 
a BSL event is very large, the light curve with lower amplification has 
little effect on that of the high amplification event, making the observed 
BSL event light curve difficult to distinguish from a SSL event one.
However, we find that the fraction of BSL events with similar source star 
amplifications is still substantial ($\sim 8\%$ for $\delta\beta\leq 0.1$),
and thus the very low detection rate of BSL events cannot be explained by 
this reason alone.
We find that the light curves of an important fraction of BSL events are 
additionally confused with SSL events due to various other reasons.
These reasons include the large brightness differences between source stars, 
small source star separations, and the imitation by BSL events of SSL events 
with longer timescales and larger impact parameters.

\acknowledgements
We would like to thank M. Everett for making helpful comments.
This research is supported by a Korean Science and Engineering
Foundation (KOSEF) grant 981-0203-010-1.

\clearpage

\begin{center}
\bigskip
\bigskip
\centerline{\small {TABLE 1}}
\smallskip
\centerline{\small{\sc The Distribution Detectable of BSL Events 
with Similar Amplifications}}
\smallskip
\begin{tabular}{ccccccccccccc}
\hline
\hline
\multicolumn{1}{c}{} &
\multicolumn{12}{c}{stellar type} \\  
\cline{2-13}
\multicolumn{1}{c}{$\delta m$} &
\multicolumn{1}{c}{} &
\multicolumn{3}{c}{A stars} &
\multicolumn{1}{c}{} &
\multicolumn{3}{c}{B stars} &
\multicolumn{1}{c}{} &
\multicolumn{3}{c}{G stars} \\  
\cline{2-13}
\multicolumn{1}{c}{} &
\multicolumn{1}{c}{} &
\multicolumn{1}{c}{$P_{3\sigma}$ } &
\multicolumn{1}{c}{$f(\delta m)$ } &
\multicolumn{1}{c}{$f_{3\sigma}$} &
\multicolumn{1}{c}{} &
\multicolumn{1}{c}{$P_{3\sigma}$ } &
\multicolumn{1}{c}{$f(\delta m)$ } &
\multicolumn{1}{c}{$f_{3\sigma}$} &
\multicolumn{1}{c}{} &
\multicolumn{1}{c}{$P_{3\sigma}$ } &
\multicolumn{1}{c}{$f(\delta m)$ } &
\multicolumn{1}{c}{$f_{3\sigma}$} \\
\multicolumn{1}{c}{(mag)} &
\multicolumn{1}{c}{} &
\multicolumn{1}{c}{($\%$)} &
\multicolumn{1}{c}{($\%$)} &
\multicolumn{1}{c}{($\%$)} &
\multicolumn{1}{c}{} &
\multicolumn{1}{c}{($\%$)} &
\multicolumn{1}{c}{($\%$)} &
\multicolumn{1}{c}{($\%$)} &
\multicolumn{1}{c}{} &
\multicolumn{1}{c}{($\%$)} &
\multicolumn{1}{c}{($\%$)} &
\multicolumn{1}{c}{($\%$)} \\
\hline
8.0   & & 0.0  & 6   & 0.0 & & 0.0   & 39  & 0.0 & & 0.0   & 25  & 0.0 \\
4.2   & & 0.0  & 29  & 0.0 & & 0.0   & 22  & 0.0 & & 0.0   & 23  & 0.0 \\
2.5   & & 6.5  & 26  & 1.7 & & 6.5   & 29  & 1.9 & & 6.5   & 37  & 2.4 \\
0.7   & & 10.7 & 39  & 4.2 & & 10.7  & 10  & 1.1 & & 10.7  & 15  & 1.6 \\
total & &      &     & 5.9 & &       &     & 3.0 & &       &     & 4.0 \\
\hline
\end{tabular}
\end{center}
\smallskip
\noindent
{\footnotesize \qquad NOTE.--- Here $P_{3\sigma}(\delta m)$ represents
the mean probability for distinguishing between a SSL event and a BSL event 
with source star brightness difference $\delta m$.  The values of 
$P_{3\sigma}(\delta m)$ are averaged over the entire
range of source star separations and impact parameters.
To show the BSL event detectability for similar source star amplifications,
the difference in impact parameters is restricted to be less than 0.1.
Then the distribution of detectable BSL events as a function of $\delta m$ 
is obtained by $f_{3\sigma}(\delta m) = P_{3\sigma}(\delta m)f(\delta m)$,
where $f(\delta m)$ is the distribution of source star brightness differences.
One finds that even with similar source star amplifications, the fraction of 
BSL events is very small ($\lesssim 6\%$) regardless of source
star spectral types.
} 
\clearpage

\postscript{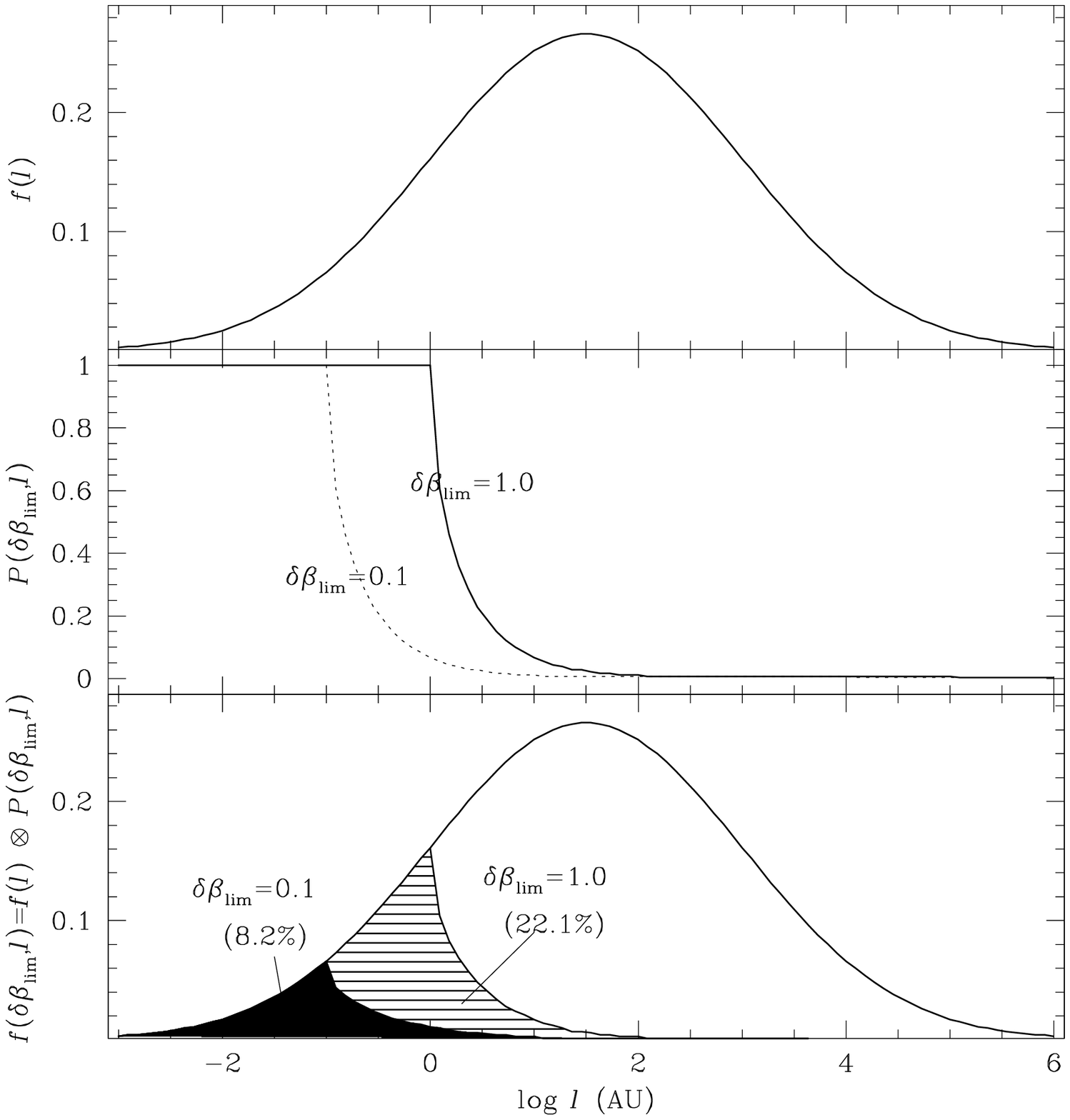}{0.95}
\noindent
{\footnotesize {\bf Figure 1:}\
Upper panel: The model distribution of binary source star separations, 
             $f(\ell)$.
Middle panel: The mean probability of BSL events having a difference in 
	      impact parameters less than a limiting value of 
	      $\delta\beta_{\rm lim}$ as a function of source star 
	      separation $\ell$, $P(\delta\beta_{\rm lim},\ell)$.
	      Here we present $P(\delta\beta_{\rm lim},\ell)$ for the two
	      cases $\delta\beta_{\rm lim}=0.1$ and 1.0.
Lower panel: The distribution of BSL events with the impact parameter
	     difference less than $\delta\beta$ as a function of $\ell$,
	     $f(\delta\beta_{\rm lim},\ell)$. 
We assume $r_{\rm E} = 1$ AU which typical for Galactic
bulge events with $M\sim 0.25\ M_\odot$, $D_{os}\sim 5\ {\rm kpc}$ and 
$D_{os}\sim 8\ {\rm kpc}$.
The fractions of BSL events with $\delta\beta$ less than 0.1 and 1.0 
are $8.2\%$ and $22.1\%$, respectively.
}\clearpage

\end{document}